\documentclass[epj,final]{svjour}

\usepackage{latexsym}
\usepackage{url}
\usepackage{amsfonts}
\usepackage{tikz}
\usetikzlibrary{decorations.pathmorphing}
\usetikzlibrary{decorations.pathmorphing,calc}
\usepackage{amsmath, amssymb}
\RequirePackage{graphicx}

\begin{document}
\title{On the origin of cosmic web}
\author{V.G.Gurzadyan\inst{1,2}, N.N.Fimin\inst{3}, V.M.Chechetkin\inst{3}, 
}                     
%
%
\institute{Center for Cosmology and Astrophysics, Alikhanian National 
Laboratory and Yerevan State University, Yerevan, Armenia \and
SIA, Sapienza Universita di Roma, Rome, Italy \and  Keldysh Institute of Applied Mathematics of RAS, Miusskaya Sq., 4, Moscow, Russia}
\date{Received: date / Revised version: date}
%

\abstract{The emergence of one and two-dimensional configurations -- Zeldovich pancakes -- progenitors of the observed filaments and clusters and groups of galaxies, is predicted by means of a developed kinetic approach in analyzing the evolution of initial density perturbations. The self--consistent gravitational interaction described by Vlasov-Poisson set of equations with branching conditions is shown to predict two--dimensional structures as of layers of increased density and voids between them, i.e. the cellular macro-structure of the Universe.  The modified potential of weak-field General Relativity is involved, which enables one to explain the Hubble tension, revealing the conceptual discrepancy in the local galactic flows and the cosmological expansion. This demonstrates the possible essential role of self-consistent gravity in the formation of the cosmic web.
}

\PACS{
      {98.80.-k}{Cosmology}   
     } 
%
\maketitle

\section{Introduction }

The evolution of the initial density perturbations, according to current views, has led to the formation of the observed large-scale
matter distribution in the Universe. The linear and non-linear phases of the evolution of perturbations are considered to be responsible
for various features, at corresponding scales \cite{ZN,Peeb,BK}. The main approaches in those studies include hydrodynamical, kinetic, 
N--body, with substantial numerical simulations.

Zeldovich's \cite{Z} pancake theory is predicting the formation of structures of particular role, one and two--dimensional 
configurations  (``pancakes'') as progenitors of the commonly observed clusters and groups of galaxies \cite{DLN,Arn,SZ,SS,Hid,Br,Feld}.  
Zeldovich approximation is an
efficient methodical tool for broader scope of physical problems \cite{Arn}.
The evolution of inhomogeneities ceases to be linear at stage when General Relativity (GR) effects become insignificant and Newtonian gravity
is applicable. The structures on the scales of galaxies, galaxy clusters and voids, arise at non-linear phase of perturbations
evolution i.e. when the density contrast becomes of order $\delta \rho / \rho \simeq 1$, and when the perturbation ceases to participate 
in the cosmological expansion and can lead to formation of gravitationally bound systems.

The methodology proposed in \cite{Z,Arn} dealing with the approximate solution of the equations of non-relativistic cosmological
dynamics is based on the use of Lagrangian variables and extrapolation of the linear theory. At sufficiently large gradients of peculiar
velocities the stage of growth of perturbations leads, based on general considerations, to a predominant
compression along one of the directions (less likely, along two directions) with the formation of flattened structures, the Zeldovich
pancakes. Mutual crossing of pancakes then leads to the creation of the web structure, so the clusters of galaxies are formed in the
high density regions, surrounding the voids.

The qualitative mechanism leading to the formation of pancakes is apparently simple, and the corresponding mathematical formalism quite
adequately describes the whole range of problems associated with the emergence of the $2$--dimensional structures \cite{Arn}.
At the same time, an ambiguity does remain determined with the initial conditions for that mechanism. Namely, a priori the presence of
perturbations of certain level of coherence is required. Within the hydrodynamic formalism it appears, for example, as a consequence of the
specific  Cauchy conditions for an evolutionary problem or as effects of periodic ``overturning'' of a propagating density wave 
(the origin of which have to be sought at previous cosmological epochs).
However, as a generating perturbation at the hydrodynamical level, that mechanism is quite sufficient for
consideration of the properties of a kinetic system with a self-consistent gravitational interaction of the particles of the system.

In our analysis, the Newtonian cosmology, known since McCrea-Milne \cite{MM} and its efficiency outlined in \cite{Z81}, is the framework, with important aspects, however. 
Namely, the generalized potential derived in \cite{Gurz1}, is used, satisfying Newton's basic condition of sphere--point mass gravity's 
equivalency.
That function includes a term corresponding to the cosmological constant term in relativistic Friedmannian equations and as 
weak--field limit of General Relativity enables one to describe the dynamics
of galaxy groups and clusters, the Hubble tension \cite{Gurz2,Gurz3,Gurz5,Gurz6,Gurz7,Gurz8}.

Thus, our novel kinetic description of the formation of web structure of the Universe within the framework
of non-relativistic gravitating model reveals the appearance of $2$--dimensional structures as of layers of increased 
density and voids between them. This approach via the analysis of Vlasov-Poisson system of equations in Newtonian gravitational configurations
\cite{VFC}, is taking into account the properties of localized semi-periodic solutions of those equations, considered as adiabatically
stable in time and with branching in the configuration space depending on the degree of dispersion of the particles in terms
of velocities and effective size of the $2$--dimensional layers.

\section{Conditions for existence of stationary solutions of Vlasov--Poisson equations}

One of the ways of the study of a system of gravitating massive particles, is the analysis
of the solutions of Einstein--Vlasov equation system \cite{And}. However, in the general case, it is an extremely time--consuming problem,
and at reasonable conditions, when the considered configuration's scale is less than the particle horizon, one can consider  the dynamics
of massive particles in cosmological context using the Vlasov--Poisson equations instead.

The problem is to inquire into the stationary solution of the Liouville equation represented as a function of the integrals of 
motion of
the particles; we'll suppose basic conserved integral is energy only. We will be interested, 
under such assumption, in the appearance in the system of periodic solutions at branching 
of solutions at certain critical points. The latter, obviously, is aimed to cover the web structure of the Universe, including the
quasi--periodical voids.

Let $f({\bf x},{\bf v},t)$ be a distribution function
 of particles (with masses $m$) by velocities
${\bf v}\in R^3$ and by coordinates
${\bf x}\in R^3$ (stars, star clusters, galaxies can be considered as particles; for simplicity of the model
we will not detail the type of structural particles and, accordingly, we will not carry out separation by masses, $m_i \equiv m$).
Assume that, $f ({\bf x},{\bf v},t)$ are solutions of  Vlasov--Poisson equation
\begin{equation}
\frac{\partial f}{\partial t} + {\bf v}\frac{\partial f}{\partial {\bf x}}-
\frac{\partial\varphi}{\partial {\bf x}} \frac{\partial f}{\partial {\bf v}}=0,~~~~
\Delta \varphi ({\bf x},t) =  4\pi G m \int f ({\bf x},{\bf v},t)\:d{\bf v},
\label{1}
\end{equation}
$$
\varphi ({\bf x},t) = \int \phi({\bf x},{\bf x}')f ({\bf x}',{\bf v},t)\:d{\bf x}'d{\bf v},~~
\phi({\bf x},{\bf x}')
$$
$$
\equiv \phi_N({\bf x},{\bf x}') =\frac{-Gm}{|{\bf x}-{\bf x}'|}, 
$$
$$
{\bf F}({\bf x}) =-\nabla_{\bf x}\int \phi ({\bf x},{\bf x}')f ({\bf x}',{\bf v},t)\:d{\bf x}'d{\bf v},
$$
\noindent
where  
$\varphi ({\bf x})$ is the self--consistent gravitational potential at the point 
${\bf x}$, $F({\bf x})$ is the self--consistent force, 
$\phi_N (r)$ is the normed inter-particle Newton attraction  potential,  
$G$ is the gravitation constant.
We are interested in solutions of system Eq.(\ref{1}) in the region
${\bf x}\in D$, where $D$ is a $3$--cube (with periodic   boundary conditions)
or an arbitrary area with a smooth boundary.

Let the distribution functions  $f$ depend on the energy integral only:
$f = g (\varepsilon)\big|_{\varepsilon =  m{\bf v}^2/2 + m\varphi (\bf x)}$,
where   $g (\varepsilon)(\ge 0)\in C^1(\Omega_\varepsilon)$.
In this case the Liouville equation of Eq.({\ref{1}}) is  satisfied (so left-hand-side of  Liouville equation is a consequence of 
Poisson brackets),
and the Poisson equation for gravitation potential has following form:
\begin{equation}
\Delta \varphi  = \Upsilon(\varphi),~~~ \Upsilon(\varphi)\equiv  4\pi G m \int g \big( m{\bf v}^2/2+
m\varphi (\bf x)  \big)\:d{\bf v}.
\label{2}
\end{equation}
According to \cite{Cour}, the Dirichlet boundary value problem ($\varphi|_{\partial D} = 0$)  for this nonlinear elliptic equation 
(in the domain $D$
with smooth border $\partial D$), belongs to the class of incorrectly posed ones in Courant's sense (it has more than one solution, 
for $d$--dimensional space: 
$D \subset R^d$, $d=1,2,3$). It should be noted that in this case, a similar problem for a system of charged 
particles for the dimension of the domain $d=1,2$ turns out to be posed correctly.

To eliminate the multiplicity of solutions of the equations of the gravitation potential for  the most physically important case of so-called Liouville--Gelfand equation \cite{Gelfand}, we will consider
the representation of the stationary distribution function in terms of
Maxwell--Boltzmann distribution in an external gravitational field
$g(\varepsilon)=\int_{R^d} n(\varphi({\bf x})) f_{eq}({\bf v})d{\bf v}$. The conditions for such a presentation are
 {\it is the maximum possible statistical independence} in the representation of a stationary distribution function
 \begin{equation}
f_{eq}({\bf v})=A_1 (d;\theta,m)\prod^d_{i=1}{\bar{f}}({v}_i^2),~~~
  A_1(d;\theta,m) = \big(m/(2\pi \theta) \big)^{d/2},
	 \label{3}
	\end{equation}
	$$
	{\bar{f}}({v}_i^2)=\exp\big(-mv_i^2/(2\theta)\big).
 $$
 
  Separation of variables when substituted into stationary
  equation (1) gives the kinetic temperature $\theta$ in the form of the separation constant, so that
  $n({\bf x})=A_2 \exp(-m\varphi({\bf x})/\theta)$ ($A_2 \equiv n_0$);
  the possible  movement of flow (with the mass velocity $ {\bf V}_0$), as usual, is taken into account
  by replacing ${{\bf v}} \to ({\bf v} - {\bf V}_0)$. Thus, the nonlinear elliptic equation (\ref{2})
  can be written (after integrating the right-hand-side) in explicit form
  \begin{equation}
  \Delta \varphi = 4\pi G m A_2\exp\bigg( -\frac{m\varphi}{\theta} \bigg),
  \label{4}
  \end{equation}
 \noindent
 or, after renaming $\varphi = -\Phi \theta/m$,
\begin{equation}
\Delta \Phi+\lambda \exp(\Phi)=0,~~~ \lambda\equiv  4\pi G  n_0\frac{m^2}{\theta}.
 \label{5}
 \end{equation}

 In the 3-dimensional case, the total mass of the system of gravitating particles is
 \begin{equation}
M_3=\int A_2 \exp(-m\varphi/\theta)d{\bf r} > 
\label{6}
\end{equation}
$$
{\lim}_{\varphi_0\to 0}
 A_2\int^{r_{max}\to \infty}_{r_0}
 \exp(m \varphi_0/\theta)\cdot 4\pi r^2 dr \to +\infty~~ (\forall r_0>0).
$$
This would imply that, 3--dimensional structures which are a solution to the Dirichlet problem for the Liouville--Gelfand equation,
   of limited mass do not exist in Euclidean 3--dimensional space. Note, however, that, 3-dimensional structures can exist when the mass integral converges, which requires either
compactness of the support, or the corresponding decay rate of the radial solution at infinity.
In this case, however, in accordance with \cite{DF}, 
for the case $n = 3$ (in contrast to the cases $ n = 1,2 $) the Liouville-Gelfand equation does not admit
$C^2$--smooth solutions that are stable outside the compact domain. Thus, three-dimensional (compact) structures
 are not excluded in the framework of the Liouville-Gelfand model.
	
	So, we turn the attention to planar and filamentary structures, as they have the physical ability to
    develop endlessly, and their total masses remain finite. Let us show this. It is possible to carry out the classification
solutions of nonlinear elliptic equation ({\ref{2}}) reduced to form ({\ref{5}}). For the case $d = 1$, consider
the formal Dirichlet problem on the segment $x \in [-R, R] \to [-1,1]$ (we will use the dimensionless reduced units $r / R$,
moreover, we assume that the value of $R$ is very large. Then the potential is physically annihilated by $\lim_{r \to R} \Phi (r) \to 0$,
but from a point $R <\infty$),
for which there is a critical parameter $\lambda_{c} = \lambda_{c}^{(1)} \approx 0.88$, such that for $0 <\lambda <\lambda_c$.
There are two strictly positive decreasing solutions (of class $C^2 ([- 1,1])$), as the argument approaches the boundary of the region  --- stable minimal $\Phi_\lambda^{min}$ and unstable $\Phi_\lambda^{sing}$  \cite {Dup1}.
The latter is with an unboundedly growing rate
at $\lambda \to 0$ $(\forall x \in [-1,1])$, which physically means an unlimited increase in the temperature of the medium
or a decrease in the mass of the structural. 
For $\lambda \to \lambda_c^{(1)}-0$ both solutions merge into one, which cannot be extended to the region $\lambda> \lambda_0$.
 For the case $d = 2$, the situation is similar: for $\lambda \in (0,  \lambda_c^{(2)} = 2)$ there is a pair of solutions
 $\Phi_\lambda^{min}$ and
 $\Phi_\lambda^{sing}$ (non--minimal unstable), but singular/non--minimal solution
 grows unboundedly only if $\lambda \to 0+$ jointly tends
 and $r \to 0$; for $\lambda \to 2$, these solutions also merge into one, which cannot be continued as $\lambda \to 2+$
 grows (see, e.g. \cite {Dup2,Dup2aaa}).
 For the case $d = 3$, there is a critical value of the parameter $\lambda_c = \lambda_c^{(3)} \approx 3.32$,
 such that there are 4 regions in the plane
 $(\lambda, \| \Phi_\lambda \|)$ allocated by
 values of the parameter $\lambda$: 1) the domain $\lambda \in (0, \lambda_c)$, where there is a
 finite number of $C^2$--smooth solutions,
 2) for $\lambda=2$ the number of solutions is infinite, 3) for $\lambda = \lambda_c^{(3)}$ there is only one solution,
 4) for $\lambda> \lambda_c^{(3)}$
 no solutions exist \cite{Dup3} (in the specified set of alternatives
 only one minimal (regular) from each set will be a stable solution).

 From a physical point of view, the existence of two types of resonances in the parameters is possible: in the first case
 the ratio $m/\theta$ (velocity dispersion of particles) forms a locally finite number of positive and radially decreasing to
 the boundary of solutions that
 can be identified with radial waves of various densities; in the second case, the presence of an infinite number of solutions
 leads to randomization of the distribution of particles in the system, since the stationary solution of the Vlasov system will be
 determined
 superposition of an infinite number of potentials. However, these physical considerations are questionable from the point of view
 of mathematical
 interpretation of the possible existence of 3--dimensional structures.

   For a $2$--dimensional system of particles, the total mass is
 $M_2 [\lambda_c] = \int^\infty_0 \int^{2\pi}_0 A_2 \big( 4/(1+r^2)^2   \big)r d\vartheta dr< \infty$
   where, for example, the value of the potential at the critical point $\lambda_c^{(2)} =2$ is chosen.
   The use for estimates of the total mass of minimal and non--minimal solutions for arbitrary
   subcritical values of the parameter ($\forall \lambda \in (0 +,2)$) gives the same finite asymptotic. Here
   we used the explicit form of solutions of the Liouville equation in the 2--dimensional case and the degeneration of two
   solutions into one at the critical point
    \begin{equation}
	\Phi_\lambda^{min/sing} =\ln\big(
    8\zeta_{min/sing}\big(   1+ \lambda \zeta_{min/sing}r^2  \big)^{-1}
    \big), 
		\label{7}
		\end{equation}
		$$
		\zeta_{min/sing}=\lambda^{-2}(4-\lambda \pm \sqrt{16-8\lambda}).
	$$
    
    Likewise, we can consider the one--dimensional case. However, for clarity of understanding, it is necessary to stipulate
   statement of the boundary value problem (verifying that of similar problem in \cite{Vlasov1}): according to the Gidas--Ni--Nirenberg 
   theorem \cite{Gidas},
   each solution of the Liouville--Gelfand equation must be radially symmetric and decreasing. 
	Therefore, the boundary conditions
   should be formulated (to exclude a non-minimal solution with an unboundedly growing norm at $\lambda \to 0$)
   to the segment $r \in [0, \tilde {r}_{max}|_{r \to \tilde {r} = r / R} = 1] $: $\Phi'(0) = \Phi (0) = 0$; it
   should be noted that, the interpretation of taking into account only the singular
   solution is a nontrivial problem of significant physical interest.

	By solving such a problem we arrive at the function
 \begin{equation}
\Phi (r) =\ln\bigg( -\frac{1}{2}(c_1^2 \lambda)^{-1} \big({\rm tanh}^2\big(  \frac{1}{2}  (r/c_1-1+c_2/c_1)  \big)-1\big) \bigg),
\label{8}
\end{equation}
 where arbitrary constants $c_1, c_2$ are determined from the boundary conditions
  ($c_1 = \pm 1 / \sqrt {2 \lambda} $, $c_2 =0$. There also exists
  a set of constants $c_{1,2}\in \mathbb{C}$, but its physical meaning is unclear, perhaps the
  formulation of the problem for the potential acquires periodicity.
	The condition of the finiteness of the mass of 1--dimensional structure
  as a solution to the boundary value problem for the Liouville equation
   \begin{equation}
	M_1[\lambda]=\int_{0}^\infty \exp\big( \ln\big(  \big[[\lambda c_1^2]^{-1}{\rm tanh}^2[x/c_1] -1\big]   \big)\big)   dx =
    \sqrt{2/\lambda}<\infty.
	\label{9}
    \end{equation}
Thus, $\Phi (r) = \ln(1 - {\rm tanh}^2 (r / M_1))$.
For the particle density, we have, in accordance with the above formula,
 $n({r}) \big|_{d = 1} = A_2 \exp (\Phi ({r})) \big|_{d = 1} \propto {\rm sech}^2 (r / M_1)$.

 Let us analyze  the 2-dimensional case.
 For a cylindrical (not necessarily hollow) structure, the solutions to the Liouville equation
 with boundary conditions similar to the $1$--dimensional case and the normalization condition in the secant (normal boundary) plane
$\int^\infty_0 \exp\big(\Phi(r)\big)2\pi r dr=M_2[\lambda]<\infty$
have different asymptotic for $r \to 0$:
  non--minimal solution in this case grows unboundedly, so we exclude it from consideration, again by an additional condition in the center.

 For a minimal (regular) solution, we obtain (for an arbitrary admissible $0 <\lambda \le 2$)
\begin{equation}
\Phi_\lambda^{min} (r) =\ln\bigg(   \frac{(32-8\lambda-8\sqrt{16-8\lambda})\lambda^{-2}}{\big(
1+\lambda^{-1}(4-\lambda-\sqrt{16-8\lambda}) r^2
\big)^2}     \bigg),
\label{10}
\end{equation}
\noindent
and $M_2[\lambda] = 8 \pi / \lambda$.

Therefore, we finally obtain $\Phi (r)= -2\ln\big(   1+ \pi^2r^2/M_2^2[\lambda]     \big)$,
and $n({\bf r})=A_2\big(   1+ \pi^2r^2/M_2^2[\lambda]     \big)^{-2}$.

Thus, cylindrical two--dimensional structures in each normal section have a finite mass, and in the presence of their
finite length, the final total mass formed due to the
of regular solution of the Liouville--Gelfand equation is selected by the condition of exclusion of the singular solution
in the center.

 The inclusion in the consideration of the singular branch of the solution of the Liouville-Gelfand equation is taken into account below, in Section 3, with an example of the transition to Cartesian coordinates in the mentioned equation. This is due to the fact that taking into account the singular solution generates ambiguity, which can be eliminated due to the translational invariance of the solution: instead of one structure considering a sequence of periodically repeating structures. Moreover, for $ j > 2 $ the sequence of bifurcations of the solution (see Fig. 1) leads to the instability of the original singular solution (with the subdivision of the scale of linear periodicity). Therefore, we consider only the stable non-minimal branch of the solution of the Liouville-Gelfand equation \cite{Du}.
We also note that, the condition for annulation of the derivative of the potential at the center of the circle (on the plane) or at the axis of the cylinder, coincides with the condition of excluding of a singular unstable solution; a similar setting is discussed in \cite{ZBS}.

Under certain conditions, the aforementioned equation admits linearization near the degeneration point of the exponential
non-linearity to constant $m\varphi/\theta \to 0$. This, for example, is possible when taking into account the rotation of
the system (solid in the case of small
angular velocities or differential in a more general case)
or taking into account the repulsive interaction on a cosmological scale with the cosmological term \cite{Gurz1}.

\section{Branching solutions of Liouville--Gelfand equation and criteria for macrostructure emergence}

Let us consider the possibility of the emergence of structures in a system of gravitating particles under cosmological conditions.
It is natural to assume that, the factors determining these processes are the properties of the dynamics of a system of particles
on a large scale in the later cosmological era. The main model, quite successful in terms of descriptions of such evolutionary dynamics
(with the conditions of uniformity of the density of distribution of particles in space and proportionality of their velocities
coordinates), as mentioned above is the non-relativistic McCrea--Milne model \cite{MM}, which at the considered distance scales 
is compatible to the relativistic Friedmannian model.

We will demonstrate the possibility of direct modification of the Liouville-Gelfand equation by 
additional terms corresponding e.g. to certain physical processes, additional interactions, the rotation of the system, etc.

It is readily obtained from the Vlasov-Poisson equations, when
the gravitational potential has a quadratic form $\phi \propto | {\bf x}-{\bf x}'|^2$ as a solution of the Poisson equation with
constant right-hand-side. This corresponds to the theorem proved in \cite{Gurz1}, stating the
generalized function satisfying Newton's identity of the gravitational field of a sphere and a point mass
\begin{equation}
\widetilde{\phi}_N(r)=-\frac{Gm}{r}-\frac{c^2\Lambda}{6}r^2.
\label{11}
\end{equation}
Note, in this potential the second term corresponds to the cosmological constant term in Friedmannian equations 
(for details see  \cite{Gurz2}).
The $\widetilde{\phi}_N(r)$ function is shown to fit the dynamics of galaxies and clusters of galaxies, also to explain the
Hubble tension i.e. the discrepancy between the Hubble constant values obtained, on the one hand, with
the data of the Planck satellite on the Cosmic Microwave Background radiation, on the other hand, through the Hubble diagram for 
galaxies in the vicinity of the Local Supercluster  \cite{Gurz2,Gurz3,Gurz5,Gurz6,Gurz7,Gurz8}.  Namely, as shown in these studies, due to the smallness of the cosmological constant the second term in Eq.(11) becomes important at scales of galactic halos and larger, i.e. of the groups and clusters of galaxies, of the local Hubble flow, etc. Naturally, conclusions obtained here for the potential of Eq.(11) should be applicable for those scales, i.e. of pancakes. The issue of extensive numerical simulations regarding pancakes, with potential of Eq.(11), is arising.   

The above--described approach to the self--consistent  Vlasov--Poisson problem, hence, has also an observational link.

For the unperturbed by quadratic term of the boundary value problem for the equation (\ref {5}), it is possible to transite to the
Cartesian $2$--dimensional $(x, y)$
coordinate system that provides consideration of this equation in the band $(x \in R^1) \times (y \in [-Y, Y])$ with the periodicity
conditions
along the abscissa axis. Then it is possible to introduce consideration of the chain of periodic structures \cite{Som}, satisfying
the condition on the nonlinear term in the Liouville--Gelfand equation:
$$
\exp(\Phi) =\frac{2j^2}{\lambda}\big(  a_1 {\rm cosh}(jy)+a_2 {\rm cos}(jx) \big)^{-2},
$$
$$
a^2_1-a_2^2=1,~~a_1\ge 1,~~j = 1,2...
$$
For $| y | \gg 1$  we get $\Phi \approx \ln \big (2j^2 \lambda^{-1} a_1^{-2} \big) -2 \ln \big ({\rm cosh} (jY) \big)$, and
thus, $\Phi \to 0$ for $y = \pm Y$ for $a_1 = j \sqrt {2 / \lambda}\,{\rm sech} (jY)$ (in this case the condition $a_2 \ge $ 1
corresponds
the condition on the eigenvalues  $\lambda \le  \lambda_j= 2j^2 {\rm sech}^2(jY)$; see Fig.1 for illustration. Obviously, the question arises whether there are
periodic solutions of the Liouville--Gelfand equation for the potential perturbed by the presence of the cosmological term.

\begin{figure}
\centering 
\includegraphics[width=0.45\textwidth]{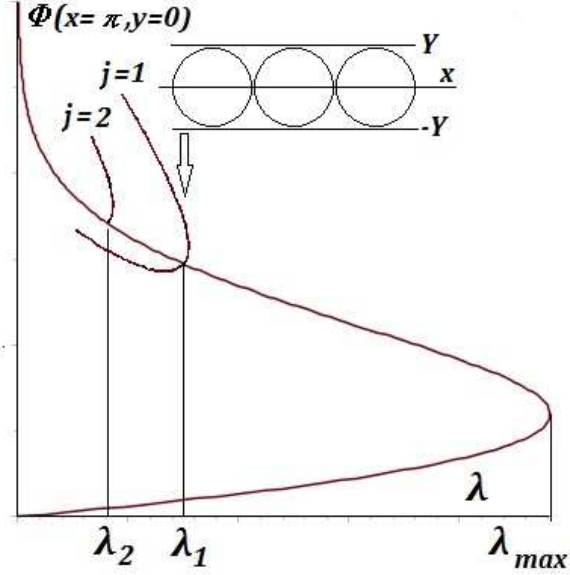}
\caption{Bifurcation diagram for the solutions of the Liouville-Gelfand equation. The periodicity along the $x$-axis arises when $a_2>0$,
the eigenvalues $\lambda_j$ correspond to new regular solutions of bifurcation branches over the original 
non minimal branch of the Gelfand problem (for hydrodynamic vortices, see \cite{Som}).}
\label{fig:lambda_hist}
\end{figure}

Note that, the separation parameter $\theta$ (formal temperature), generally speaking, does not need to be positive.
In the case $\theta<0 $, we obtain, respectively, $\lambda <0$ (for the $2$--dimensional case of the logarithmic interaction potential
we get an analogue of the Onsager vortex system, in which negative temperatures are admissible). This opens up an attractive problem,
from the point of view of the prospects, of inquiring into the solutions of the Liouville--Gelfand equation in cosmology, since one of 
the admissible solutions $\widetilde{\varphi}(r) = \ln \big (2 \eta_1^2 / |\lambda|\cdot\big(r \cdot{\rm sin} 
(\eta_2 r^{\eta_1}) \big)^{-2}\big)$ in this case is the quasi-periodic one along the radius, with decreasing amplitude.

Consider the possibility of the appearance of periodic structures in the one--dimensional case, using the linearization of the
Liouville--Gelfand equation
under conditions when it is possible to use as the inter-particle interaction potential the corresponding to
formula (\ref{11}). The equation corresponding to (\ref{4}) for the modified potential $\widetilde{\varphi} (x)$ has the form
\begin{equation}
\frac{d^2}{dx^2}\widetilde{\varphi} =\lambda\exp\bigg( -\frac{m   \widetilde{\varphi}}{\theta} - \frac{mc^2\Lambda x^2}{6\theta} \bigg),~~~
\lambda \equiv 4\pi Gm  A_2.
\label{12}
\end{equation}
\noindent
Suppose that, the expression in the exponent is small, so that it can be replaced by the first two terms of its expansion in the
vicinity of zero.
We get an ordinary differential equation of the 2nd order
\begin{equation}
\frac{d^2}{dx^2}\widetilde{\varphi}+(\lambda\alpha)\widetilde{\varphi} + (\lambda \beta x^2 -\lambda)=0, ~~~\alpha=\frac{m}{\theta},~~
\beta= \frac{m  c^2 \Lambda}{6 \theta}.
\label{13}
\end{equation}
Its solution can be written in quadratures, but it is rather cumbersome. Let us present solutions in degenerate cases:

1) if in the coefficients
the quantity $\beta x^2 $ (long--range asymptotic) prevails: 
\begin{equation}
\widetilde{\varphi}(x) =K_1 \sin(\sqrt{\alpha\lambda}x)+K_1 \cos(\sqrt{\alpha\lambda}x)-\frac{\beta(\alpha
\lambda x^2-2)}{\alpha^2\lambda},
\label{14}
\end{equation}

2) if $x \ll 1$:
\begin{equation}
\widetilde{\varphi}(x) =K_3 \sin(\sqrt{\alpha\lambda}x)+K_4 \cos(\sqrt{\alpha\lambda}x)+\frac{1}{\alpha}~~~(K_j={\rm const}_{j}).
\label{15}
\end{equation}

Thus, it can be seen that the solutions of the linearized potential equation are composition of oscillatory
ones
(with the period $T_1 = 2 \pi / \sqrt {\lambda \alpha}$, note, the inverse dependence on temperature and the direct
dependence on the density $n(0)$)
and a term that changes the growth from constant to quadratic function
in the far zone dominating over oscillations.

A natural question arises: will
the Liouville--Gelfand equation linearized in a neighborhood of the points of the solution $(\lambda, \Phi)$
 have a solution of similar behavior?
 Using the linear equation as the initial one, it is possible to construct a system of approximations to the solution of the corresponding
 nonlinear equation (in our particular case, the Liouville--Gelfand equation),
 using the Puiseau expansion in terms of the root of the small parameter in the vicinity of the known solution. Following
 \cite{Vlasov1,Vlasov2},
 consider the first approximation equation $\Delta \Phi_1 + \lambda_{*}\exp(\Phi[\lambda_{*}])\Phi_1 =0$,
 where the parameter $\lambda_{*}$ is determined by the additional condition of the finiteness of the mass
 flat layer with the density distribution law obtained above normal to the layer
  or the radial section of the cylinder (above was
 the connection between the parameter $\lambda^{(d)}$ and the finite masses $M_1, M_2$ is established). Therefore, for the case when the density changes along
 thickness of the layer, we obtain  $\exp(\Phi_1[\lambda_{*}])=1-{\rm tanh}^2(2r/\lambda_{*})$, the corresponding
 equation of the first approximation
 has the form of a second order nonlinear ODE, the general solution of which is expressed in terms of generalized hypergeometric functions
  $$
\Phi_1[\frac{1}{2}](r) =\bigg(c_1 (2{\rm cosh}(8r)-2)^{3/4} {}_2F_1 \big([3/4-
$$
$$
3\sqrt{2}/16,3/4-3\sqrt{2}/16],[1-3\sqrt{2}/8],
$$
$$
\frac{1}{2}{\rm cosh}(8r)+\frac{1}{2}\big)\sqrt{2{\rm cosh}(8r)+2}(8 {\rm cosh}^4(r) -
$$
$$
8 {\rm cosh}^2(r)+1)^{-3\sqrt{2}/8}+
$$
$$
+c_2 (2{\rm cosh}(8r)-2)^{3/4} {}_2F_1 \big([3/4+3\sqrt{2}/16,3/4+3\sqrt{2}/16],
$$
$$
[1+3\sqrt{2}/8],
$$
$$
\frac{1}{2}{\rm cosh}(8r)+\frac{1}{2}\big)\sqrt{2{\rm cosh}(8r)+2}(8 {\rm cosh}^4(r)
$$
\begin{equation}
 -8 {\rm cosh}^2(r)+1)^{3\sqrt{2}/8}\big)
({\rm sinh}{(8r)})^{-1/2}, ~~~\lambda_{*}=\frac{1}{2}.
\end{equation}

 If we consider a cylindrical sleeve, then in this case, obviously,
$\exp(\Phi[\lambda_{*}]) =\big(   1+\lambda_{*} r^2/8      \big)^{-2}$,
the general solution of the equation of the first approximation has the form
\begin{equation}
\Phi_1=c_1 \frac{\lambda_{*} r^2-8}{\lambda_{*} r^2+8}+c_2 \frac{\lambda_{*}r^2\ln(r)-8\ln(r)-16}{\lambda_{*} r^2+8}.
\end{equation}
 The equations for the second and subsequent approximations $\Phi_{j>2}$ allow us to form the Puiseau series $\sum_{k = 1} 
 \epsilon^{k/2} \Phi_k$,
 approximating the exact solution. 

It is important to outline that, the formation of a two--dimensional periodical structure
starts with a local spontaneous or external action--induced violation of isotropy of the primary particle flux i.e. at
 increase in density or mass flow rate or of the energy in allocated direction of propagation. Then, the conditions for the formation of structures are satisfied, since the boundedness of the local action ensures the convergence of the integrals
in item 2. In this case, structure formation does not occur in less energetically loaded directions, and, moreover, occurs under certain 
conditions, for example,
in the vicinity of the ordered distribution of the subsystem of more massive particles, with the formation of a secondary flow of more
light particles in the vicinity.

The formed periodical structure and particularly the low density regions, voids, can lead to observable effects regarding 
the photon beam propagation which differ from those of homogeneous matter distribution.

Namely, representing the voids as perturbations to the Friedmann-Lemaitre-Robertson-Walker  metric \cite{GK2}
\begin{equation}
ds^2 =  -(1+2\phi)\ dt^2 + (1-2\phi)\ a^2(t)\ \gamma_{mn}(x)dx^mdx^n,
\end{equation}
at perturbation $|\phi|\ll 1$, one will have for the averaged Jacobi equation of geodesics deviation
\begin{equation}
\frac{d^2\ell}{d\lambda^2}+ r\ \ell = 0,
\end{equation}
where
\begin{equation}
\lambda(z,\Omega_\Lambda,\Omega_m)
=\int_0^z\frac{d\xi}{\sqrt{\Omega_\Lambda +[1- \Omega_\Lambda + \Omega_m \xi]\ (1+\xi)^2}}
\end{equation}
and
\begin{equation}
r = -\Omega_k +2\Omega_m\delta_0,
\end{equation}
\begin{equation}
\delta_0 \equiv \frac{\delta\rho_0}{\rho_0},
\end{equation}
$\delta_0$ is the density contrast, at mean density $\rho_0$;  $\Omega_i, i=k, m, \Lambda$ are the density parameters for the curvature, 
matter and dark energy, respectively.

It follows that even at flat Universe, $\Omega_k=0$, the density contrast in the voids is able to lead to the hyperbolicity of the photon 
beams. This effect of image distortion due to semi-periodical voids have to contribute to the redshift distortion detected at galaxy 
surveys \cite{S1,S2}.

\section{Conclusions}

By means of a developed kinetic approach we show the emergence of the two--dimensional structures, as key features on the way of formation of 
the observed large scale matter distribution in the Universe. Those structures, Zeldovich pancakes \cite{Z}, have been
 predicted at the density perturbations analysis within classical or (weakly) relativistic hydrodynamics. That picture assumed certain type 
 of initial density perturbations, with the primary feature of the
 spatially separated maxima of the velocity (or momentum) distribution of the gravitating
 particles of the original system.

 The approach developed here is based on the kinetic description of evolution
 a multi-particle ensemble in a quasi--stationary state in non--relativistic regime, so that, the presence of inter-particle interaction is
 is considered on cosmological scales smaller than the particle horizon. Then, the natural technique for studying many-particle
 self-consistent dynamics becomes the formalism of the Vlasov equation, since Boltzmann collisions in this case
 have negligible role.

 The hydrodynamical technique is focused on the study of a perturbed flow in external
  fields, and the inter-particle interaction enters only in the viscosity coefficients
  and thermal conductivity, which can be influenced by other factors. In particular, at numerical simulations,
  the inclusion of the so-called grid viscosity, or the presence of a local vortex motion, essentially
  affects the mentioned coefficients. At the same time, in the Vlasov equation,
  the presence and importance of a self-consistent interaction term is indicated explicitly, and its role is
  not masked by modeling details.

In our analysis of the formation of two-dimensional structures on relevant cosmological scales, given the practically flat geometry of the 
Universe, the weak-field limit of General Relativity in involved, justifying the study of structures using the Vlasov--Poisson equations.
  In fact, we were looking for the analogue of Bernstein--Greene--Kruskal-type solutions \cite{BGK} for gravitational interaction.
  Carrying out a comparative analysis of the Poisson equation in the case of stationarity of the Vlasov equation, we obtain
  with one-dimensional mass motion of a system of particles, the appearance of semi-periodical along the spatial variable
  and parallel to the vector of the regular motion of planar structures, with
  rapid density decrease along the vector of the mass flux.

Thus, in this paper we advance a kinetic description of the formation of the cosmological cellular structure, with
voids separated by two--dimensional surfaces. The presence of those structures is determined by the quasi--oscillator equation being the 
consequence of the Poisson equation.

This approach shows that, the two-dimensional cosmological structures, the Zeldovich pancakes, can be formed not only being determined by the specific type of initial density perturbations, but also due to gravitational self--consistent interaction between the particles. This offers a new principal landscape to the formation paths of the cellular macro-structure of the Universe.

We thank A.G.Doroshkevich for valuable discussions.

\end{document}